\documentclass[twocolumn]{emulateapj}

\usepackage{xspace}
\usepackage{amsmath}
\usepackage{framed} 
\usepackage{txfonts}
\usepackage{epstopdf} 
\usepackage{color}
\usepackage{rotating}
\usepackage{natbib}
\usepackage{ulem}

\slugcomment{To be submitted to the Astrophysical Journal}

\newcommand{\etal}{{et al.~}}

\newcommand{\Msun}{\>{\rm M_{\odot}}}
\newcommand{\Lsun}{\>{\rm L_{\odot}}}

\newcommand{\beq}{\begin{equation}}
\newcommand{\eeq}{\end{equation}}

\newcommand{\rmd}{{\rm d}}

\newcommand{\msunh}{\>h^{-1}\rm M_\odot}

\newcommand{\avg}[1]{\langle #1 \rangle}

\newcommand{\drm}{{\rm d}}

\def\gtsima{$\; \buildrel > \over \sim \;$}
\def\ltsima{$\; \buildrel < \over \sim \;$}
\def\prosima{$\; \buildrel \propto \over \sim \;$}
\def\gsim{\lower.7ex\hbox{\gtsima}}
\def\lsim{\lower.7ex\hbox{\ltsima}}
\def\simgt{\lower.7ex\hbox{\gtsima}}
\def\simlt{\lower.7ex\hbox{\ltsima}}
\def\simpr{\lower.7ex\hbox{\prosima}}

\newdimen\hssize
\newdimen\hdsize
\def\rmb{{\rm b}}
\def\rmc{{\rm c}}
\def\rmd{{\rm d}}

\def\rmm{{\rm m}}

\def\rms{{\rm s}}

\def\rmx{{\rm x}}

\shorttitle{Magnitude Gap Statistics and the Conditional luminosity function}
\shortauthors{More S.}

\begin{document}
\setlength{\hbadness}{10000}


\title{Magnitude Gap statistics and the Conditional luminosity function}
\author{Surhud More \altaffilmark{1,2,3}}

\altaffiltext{1}{surhud@kicp.uchicago.edu}	   
\altaffiltext{2}{Kavli Institute for Cosmological Physics, The University of Chicago, Chicago, IL 60637 USA} 
\altaffiltext{3}{Enrico Fermi Institute, The University of Chicago, Chicago, IL 60637}


\begin{abstract}
In a recent preprint, \citet[][H12]{Hearin2012} suggest that the halo
mass-richness calibration of clusters can be improved by using the
difference in the magnitude of the brightest and the second brightest
galaxy (magnitude gap) as an additional observable. They claim that
their results are at odds with the results from
\citet[][PS12]{Paranjape2012} who show that the magnitude distribution
of the brightest and second brightest galaxies can be explained based
on order statistics of luminosities randomly sampled from the total
galaxy luminosity function. We find that a conditional luminosity
function (CLF) for galaxies which varies with halo mass, in a manner
which is consistent with existing observations, naturally leads to a
magnitude gap distribution which changes as a function of halo mass at
fixed richness, in qualitative agreement with H12.  We show that, in
general, the luminosity distribution of the brightest and the second
brightest galaxy depends upon whether the luminosities of galaxies are
drawn from the CLF or the global luminosity function. However, we also
show that the difference between the two cases is small enough to
evade detection in the small sample investigated by PS12. This shows
that the luminosity distribution is not the appropriate statistic to
distinguish between the two cases, given the small sample size. We
argue in favor of the CLF (and therefore H12) based upon its
consistency with other independent observations, such as the
kinematics of satellite galaxies, the abundance and clustering of
galaxies, and the galaxy-galaxy lensing signal from the Sloan Digital
Sky Survey. 
\end{abstract}



\section{Introduction}
\label{sec:intro}

The abundance of halos at the massive end is sensitive to variations
in cosmological parameters such as the matter density parameter
($\Omega_\rmm$) and the amplitude of the power spectrum of density
fluctuations in the Universe (characterized by $\sigma_8$). Therefore
observations of the abundance of galaxy clusters which reside in such
massive halos can be used to constrain these cosmological parameters
\citep[see e.g][]{Vikhlinin2009,Mantz2010,Rozo2010}. Additionally,
such observations can also be used to constrain the phenomenological
behavior of dark energy and test the nature of gravity through
measurements of the growth of structure, questions which are of
fundamental importance for cosmologists today \citep[see
e.g.,][]{Rapetti2010, Rapetti2012}. Photometric surveys such as the
Dark Energy Survey \citep[][DES]{Frieman2005}, the Hyper-Suprime Cam
survey \citep[][HSC]{Miyazaki2006} in the immediate future and the
Large Scale Synoptic Telescope Survey (LSST) in the near future, will
result in a large catalog of optically-selected galaxy clusters, which
can be used to answer these important questions.

Identifying the galaxy cluster observables in optical surveys which
tightly correlate with halo mass, establishing the scaling relations
between these observables and halo mass and the scatter in these
relations are all crucial steps in order to achieve the scientific
goals. It is well known that the number of cluster members (also
called richness) correlates with halo mass \citep[see
e.g,][]{Becker2007, Johnston2007, Sheldon2009}, and so does the
luminosity (or stellar mass) of the brightest (or central) galaxy
\citep{Mandelbaum2006, More2009a, Moster2010, Behroozi2010, More2011a}
or the total stellar or luminosity content in the group \citep[see
e.g.,][for results based on abundance matching]{Yang2007}. However,
these scaling relations have considerable scatter, and therefore
combining multiple observables, especially those which come without
additional observational costs, is an important task. 

Recently, \citet{Hearin2012} suggested that at fixed richness, the
magnitude difference (also called magnitude gap or equivalently the
luminosity ratio) between the brightest and the second brightest
galaxy, contains information about halo mass. By using a simple
subhalo abundance matching prescription they showed that at fixed
richness, the small (large) magnitude gap systems are expected to
preferentially reside in less (more) massive halos. To test the
proposition with real data, the authors used the galaxy group catalog
of \citet{Berlind2006} and showed that at fixed velocity dispersion
(proxy for halo mass), the average richness of small magnitude gap
systems is significantly larger than the average richness of the large
magnitude gap systems, and the difference is larger than that expected
if the luminosities of galaxies in every group were drawn randomly
from the galaxy luminosity function.

Theoretically, it is expected that dynamical friction causes brighter
satellites in massive halos to merge with the central galaxy,
increasing the magnitude gap between the brightest satellite and the
central galaxy. This suggests that central galaxies are expected to be
special, they occupy the deepest portion of the potential well, where
they grow by feeding on the satellites that are dragged to the center
of halos by dynamical friction. Whether this is indeed the case, or
whether the luminosity of the central galaxies is just a matter of
chance, is a matter which can be settled with observations.

In \citet{Paranjape2012}, the authors examine this question by looking
at the luminosity distribution of the brightest and the second
brightest galaxy and the magnitude gap between the two using the group
catalog of \citet{Berlind2006}. They find that the luminosity
distributions are consistent with the distribution of the brightest
and the second brightest of $N$ random draws from the galaxy
luminosity function, where $N$ is the richness of a given group. On
face value, this would imply that there is nothing special about the
brightest galaxy in a given group, and that it is just a matter of
chance that any galaxy becomes the brightest in a given group. These
results imply that the magnitude gap should not contain any more
information about the halo mass, than that contained in the richness,
in apparent contrast with the results from \citet{Hearin2012}, which
are based on the same group catalog.

In this paper, we attempt to clarify this issue, by predicting the
magnitude gap based upon the conditional luminosity function (CLF),
which describes the halo occupation distribution of galaxies in a halo
of given mass. The CLF and its variation with halo mass has been
calibrated using a wide variety of observations such as the abundance
of galaxies, their clustering and the galaxy-galaxy lensing signal
measured from the Sloan digital sky survey \citep[][SDSS]{York2000}.
We show that if galaxies occupy halos according to the CLF, it is
natural to expect that the magnitude gap depends upon the halo mass at
fixed richness. We also show that the luminosity distributions of the
brightest and the second brightest galaxies are predicted to be
different from those obtained by random draws from the luminosity
function. However, detecting this small difference just using the
luminosity distribution will require sample sizes which are larger
than the one used by \citet{Paranjape2012}.

We also note that in their paper, \citet{Paranjape2012} investigate
the luminosity-weighted marked correlation function and show that its
radial dependence implies that the luminosities of the brightest
galaxies are not a matter a chance. They conclude that their results
falsify the hypothesis that the luminosities of the brightest and the
second brightest galaxy are drawn from the global luminosity function
(i.e., without any dependence on halo mass or environment) and that
the luminosity distribution alone is not an appropriate discriminant
to investigate this issue. In this paper, our results based on the CLF
will strengthen this argument.

This paper is organized as follows. In Section~\ref{sec:clf}, we
describe the CLF framework and give analytical expressions for the
magnitude gap distribution based upon the CLF. In
Section~\ref{sec:sims}, we show the magnitude gap distribution from
Monte Carlo simulations based on the CLF and compare the results to
the analytical expression presented in Section~\ref{sec:clf}. We also
investigate the dependence of the magnitude gap upon the richness in a
group and the assumed CLF parameterisation. In
Section~\ref{sec:lumdist}, we construct mock galaxy catalogs based
upon galaxy luminosities sampled from (a) the CLF and (b) the overall
galaxy luminosity function and compare the luminosity distributions of
the brightest and the second brightest galaxy in these two catalogs.
Finally, we summarize our results in Section~\ref{sec:summary}.

For the purposes of this paper, we adopt the following convention. We
refer to galaxies as centrals (satellites), if they are drawn from the
CLF which is specific to the central (satellite) galaxies (see
Eqs.~\ref{phi_c} and \ref{phi_s}). As our fiducial model, we assume
that central galaxies are also the brightest galaxies in the halo.
Therefore, in the fiducial case, the magnitude gap is the difference
in magnitudes between the central galaxy and the brightest satellite.
However, we will also investigate cases, when the satellites are
allowed to be brighter than the central galaxy \citep[see][for
observational evidence of such a possibility]{Skibba2011}.

\begin{figure*}
\centering
\includegraphics[scale=0.7]{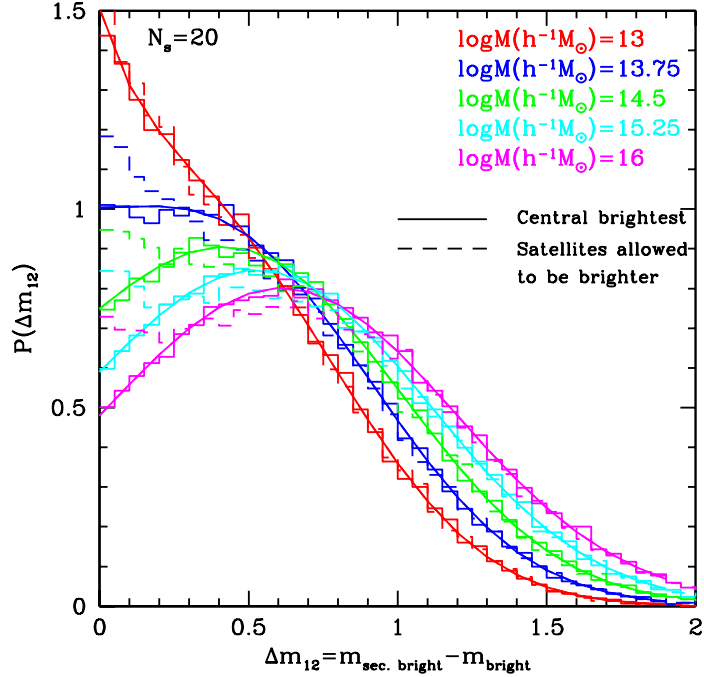} 
\caption{The distribution of the difference in magnitudes between the
brightest and the second brightest galaxy predicted by simulations in
which we populate galaxies in halos of different mass according to the
conditional luminosity function. The solid histograms show the
distribution of magnitude gaps in halos of different mass (shown using
different colors) for our fiducial model in which we assume that the
central galaxy (defined to be drawn from the central CLF) is the
brightest in the halo. The solid curves show the analytical
prediction based on Eq.~\ref{eq:pred}. The dashed histograms show the
corresponding result for the case when we allow satellites to be
brighter than the central galaxy.}
\label{fig:maggap}
\end{figure*}
\section{Conditional luminosity function}
\label{sec:clf}

The conditional luminosity function, denoted by $\Phi(L|M)$, is
defined to be the average number of galaxies of luminosities $L\pm
\drm L/2$ that reside in a halo of mass $M$ \citep{Yang2003}. The
average number of galaxies in a given halo of mass $M$ can be found by
simply integrating the CLF over the luminosities of interest, e.g.,
the average number of galaxies with luminosities between $L_{\rm min}$
and $L_{\rm max}$ that reside in a halo of mass $M$ is given by
\begin{equation}
    \avg{N}_M(L_{\rm min},L_{\rm max}) = \int_{L_{\rm min}}^{L_{\rm
    max}} \Phi(L|M)
    \drm L \,.
    \label{eq:ngal}
\end{equation}
For convenience, the CLF is divided in to a central galaxy component
($\Phi_\rmc[L|M]$) and a satellite galaxy component
($\Phi_\rms[L|M]$). 

We assume that the distribution $\Phi_\rmc(L|M)$ is described by a
lognormal distribution with a scatter, $\sigma_\rmc$, that is
independent of halo mass, consistent with the findings from studies of
satellite kinematics \citep{More2009b, More2009a, More2011a} and
galaxy group catalogs \citep{Yang2009},
\begin{equation}\label{phi_c}
\Phi_\rmc(L|M) \,{\rmd}L = {\log\, e \over {\sqrt{2\pi} \, \sigma_\rmc}} 
{\rm exp}\left[- { {(\log L  -\log L_\rmc )^2 } \over 2\,\sigma_\rmc^2} \right]\,
{\rmd L \over L}\,.
\end{equation}
The dependence of the logarithmic mean luminosity, $\log
\tilde{L}_\rmc$, on halo mass is given by
\begin{equation}
\log \tilde{L}_\rmc(M)=\log \left[ L_0
\frac{(M/M_1)^{\gamma_1}}{\left[1+(M/M_1)\right]^{\gamma_1-\gamma_2}}
\right]\,.
\end{equation}
Four parameters are required to describe this dependence; two
normalization parameters, $L_0$ and $M_1$ and two parameters
$\gamma_1$ and $\gamma_2$ that describe the slope of the
$\tilde{L}_\rmc(M)$ relation at the low mass end and the high mass
end, respectively.

The satellite CLF, $\Phi_\rms(L|M)$ is assumed to be a
Schechter-like function, 
\begin{equation}
\Phi_\rms(L|M) \drm L=\Phi_\rms^*\left(\frac{L}{L_*}\right)^{\alpha_\rms}\,
\exp\left[-\left( \frac{L}{L_*} \right)^p \right] \,\frac{\drm L}{L_*}.
\label{phi_s}
\end{equation}
Here $L_*(M)$ determines the knee of the satellite CLF and is assumed
to be a factor $f_\rms$ times fainter than $\tilde{L}_\rmc(M)$.
Motivated by results from the SDSS group catalog of \citet{Yang2008a},
we set $f_\rms = 0.562$ \citep[see also][]{Reddick2012}, $p=2$, and
assume that the faint-end slope of the satellite CLF is independent of
halo mass.  The logarithm of the normalization, $\Phi_\rms^*$ is
assumed to have a quadratic dependence on $\log M$ described by three
free parameters, $b_0$, $b_1$ and $b_2$;
\begin{equation}
\log \Phi_\rms^*=b_0+b_1\,(\log M-12)+b_2\,(\log M-12)^2\,.
\end{equation}
Note that this functional form does not have a physical motivation; it
merely provides an adequate description of the results obtained by
\citet{Yang2008a} from the SDSS galaxy group catalog. The parameters
of the conditional luminosity function and their variation with halo
mass can be constrained by using observations of the abundance, the
clustering and the galaxy-galaxy lensing signal measured from the
Sloan Digital Sky Survey \citep{More2012fisher,More2012,Cacciato2012}.
In what follows, we will use the following values for the CLF
parameters: $L_0 = 10^{9.95} h^{-2} \Lsun$, $M_1 = 10^{11.27}
h^{-1}\Msun$, $\sigma_\rmc = 0.156$, $\gamma_1=2.94$,
$\gamma_2=0.244$, $\alpha_s=-1.17$, $b_0=-1.42$, $b_1=1.82$, and
$b_2=-0.30$, consistent with the results presented in
\citet{Cacciato2012}.  

If the luminosities of galaxies in a halo are drawn in an uncorrelated
fashion, the probability that a halo of mass $M$ and richness $N$ has
a magnitude gap, $\Delta m$, or equivalently the luminosity ratio,
$f_L$, between the brightest satellite galaxy and the central galaxy
in a halo of mass $M$ is then given by\footnote{Note that our
expression differs from \citet{Paranjape2012} because in our case
the central galaxy luminosity is assumed to be sampled from a
probability distribution which differs from the distribution from
which the satellites are sampled from.}
\begin{eqnarray}
    P(f_L|N,M)=
    (N-1)\int_{L_{\rm min}}^{\infty}&& \drm L'\,
    \,P_\rms(L'|M)\,P_\rmc\left(L'/f_L|M\right)\, \nonumber \\
    &&\times\left[ P_\rms(<L'|M) \right]^{(N-2)}\,.
    \label{eq:pred}
\end{eqnarray}
Here, the probabilities $P_\rmx(L'|M)$ and $P_\rmx(<L'|M)$ are defined
such that
\begin{eqnarray}
P_\rmx(L'|M)=\frac{\Phi_\rmx(L'|M)}{\avg{N_\rmx}_M(L_{\rm
min},L_{\rm max})} \\
P_\rmx(<L'|M)=\frac{\avg{N_\rmx}_M(L_{\rm
min},L')}{\avg{N_\rmx}_M(L_{\rm
min},L_{\rm max})}
\end{eqnarray}
where the symbol $\rmx$ can either stand for central ($\rmc$) or
satellite ($\rms$).  The quantities $\avg{N_\rmx}_M$ in the relevant
luminosity intervals can be obtained by replacing $\Phi(L|M)$ by
$\Phi_\rmx(L|M)$ inside the integral in Eq.~(\ref{eq:ngal}).  For
central galaxies we choose $L_{\rm max}=\infty$. In our model, we
assume that the central galaxies are always the brightest in the halo.
Therefore, in the case of satellites, we use the luminosity of the
central under consideration as the upper limit, i.e., $L_{\rm
max}=L'/f_L$.

The integrals for $\avg{N_\rmc}_M$ and $\avg{N_\rms}_M$ can be written
in terms of the complementary error function and the incomplete gamma
function, respectively, such that
\begin{eqnarray}
\avg{N_\rmc}_M(L_1,L_2)= \frac{1}{2}&&\left[ 
{\rm erfc}\left(\frac{\log L_1-\log
\tilde{L}_\rmc}{\sqrt{2}\sigma_\rmc} \right) \right. \nonumber \\
&& \left. -{\rm erfc}\left(\frac{\log L_2-\log
\tilde{L}_\rmc}{\sqrt{2}\sigma_\rmc}\right)
\right] \\
\avg{N_\rms}_M(L_1,L_2)= \frac{\Phi_*}{p}&&\left(
    \Gamma\left[\frac{\alpha_\rms+1}{p},
    \left(\frac{L_1}{L_*}\right)^p\right] \right. \nonumber \\
    &&-\left. \Gamma\left[\frac{\alpha_\rms+1}{p},
        \left(\frac{L_2}{L_*}\right)^p\right] \right) \,.
\end{eqnarray}

The probability of a halo to have a certain mass, given its richness
and the magnitude gap can be obtained from Eq.~\ref{eq:pred} and the
Bayes' theorem,
\begin{equation}
    P(M|f_L,N) = \frac{P(f_L|M,N)P(M|N)}{P(f_L|N)} \,,
\end{equation}
and as expected it depends upon the mass-richness relation via the
probability distribution $P(M|N)$. The probability distribution within
a given bin of richness $[N_1,N_2]$ is given by
\begin{equation}
    P(M|f_L,N_1<N<N_2) = \sum_{N=N_1}^{N_2}
    \frac{P(f_L|M,N)P(N|M)P(M)}{P(f_L|N)} \,.
\end{equation}
Finally, the distribution of the magnitude gap at fixed halo mass
(without regard to the richness) is given by
\begin{equation}
P(f_L|M) = \sum_{N=2}^{\infty} P(f_L|N,M) P(N|M) \,.
\label{eq:vdb07}
\end{equation}

In what follows, we will also investigate the effect of allowing
satellite galaxies to be brighter than the central galaxies in their
halo. The analytical expressions for predicting the magnitude gap
distribution in this case are presented in the appendix.

\section{Results from Simulated Sample}
\label{sec:sims}

We now demonstrate explicitly that for fixed richness, the CLF
predicts that the magnitude gap in a given group of galaxies depends
upon the mass of the halo in which these galaxies reside. The CLF
varies with halo mass and therefore it is not surprising that this
indeed is the case. For this purpose, we generate Monte-Carlo samples
of galaxies that populate halos according to the CLF in the following
manner.

For a halo of given mass, we first draw the luminosity of its central
galaxy from $\Phi_{\rm cen}(L|M)$, given by Eq.~(\ref{phi_c}). In
order to avoid the existing correlation between halo mass and richness
affecting our conclusions, we fix the number of satellites $N_{\rm
sat}=20$. For each of the $N_{\rm sat}$ satellites, we then draw a
luminosity from the satellite CLF $\Phi_{\rm sat}(L|M)$, given by
Eq.~(\ref{phi_s}). While drawing the satellite luminosities, we adopt
a luminosity threshold, $L_{\rm min}$, corresponding to $^{0.1}M_r -
5\log h = -19$ (here $^{0.1}M_r$ indicates the SDSS $r$-band
magnitude, $K$-corrected to $z=0.1$; see Blanton \etal 2003). As
mentioned before, we also assume that the satellites are always
fainter than the central galaxy drawn for a given halo. 

The resultant distribution of the magnitude gaps is shown in
Fig.~\ref{fig:maggap} using a solid histogram for a wide range of halo
masses. It can be clearly seen that the distribution of the magnitude
gaps depends upon halo mass. We use Eq.~\ref{eq:pred} to predict this
distribution analytically and compare it to the results from our
simulations. The result of this analytical calculation are shown as
solid curves in Fig.~\ref{fig:maggap} which agrees well with the
magnitude gap distribution from our simulations.

For low mass halos, the distribution of magnitude gaps is peaked at
zero. However, this peak shifts away from zero as we move to larger
halo masses. This figure establishes that if galaxies populate halos
according to the conditional luminosity function (which is supported
by several observations such as galaxy group catalog, and the
observations of abundance clustering and galaxy galaxy lensing from
SDSS), the magnitude gap should have more information about the halo
mass, in addition to that conveyed by richness alone. At fixed
richness, higher mass halos tend to have larger magnitude gaps, in
agreement with \citet{Hearin2012}. Our result that the magnitude gap
distribution for low mass halos is peaked at zero, and shifts to
larger magnitude gaps for larger mass halos, may appear to be exactly
opposite of the result presented in \citet{vdb2007}. However, note
that the magnitude gap distributions we present are at {\it fixed}
richness and halo mass ($P[f_L|N,M]$), while the magnitude gap
distributions shown in the different panels in fig.  5 of
\citet{vdb2007} correspond to groups with {\it varying} richness
(thus corresponding to $P(f_L|M)$, see Eq.~\ref{eq:vdb07}), due to the
underlying mass-richness relation. We will shortly consider the effect
of changing richness on the magnitude gap distribution.

\begin{figure*}
\centering
\includegraphics[scale=0.8]{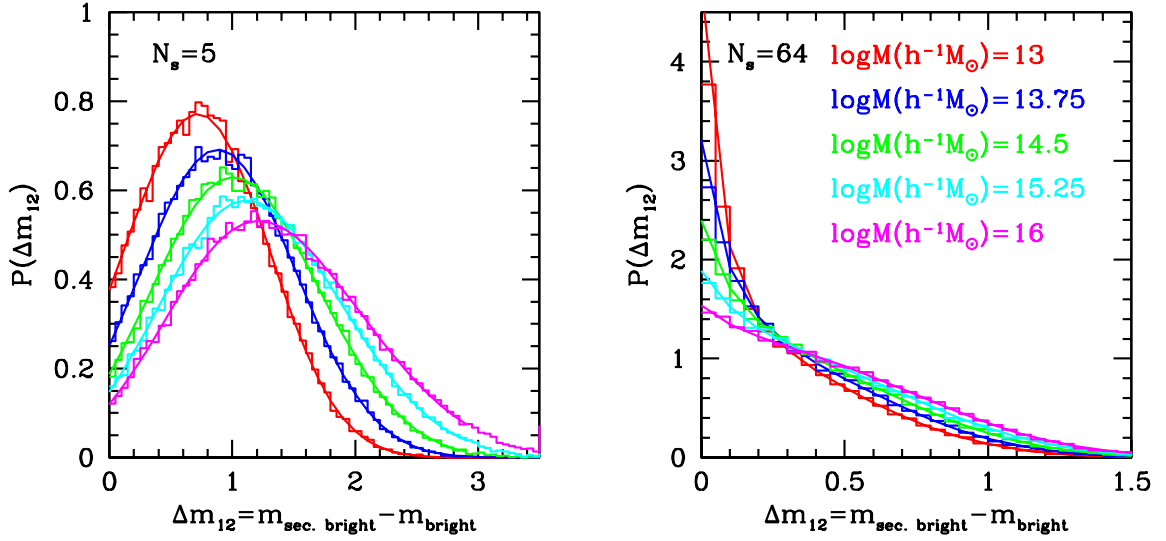} 
\caption{Distribution of magnitude gaps (similar to
Fig.~\ref{fig:maggap}) for the case when the number of satellites is
equal to 5 and 64 is shown in the left and right hand panel,
respectively, and assuming that the central galaxy is the brightest in
the halo.
}
\label{fig:maggap_varyn}
\vspace{0.2cm}
\end{figure*}

\begin{figure}
\centering
\includegraphics[scale=0.8]{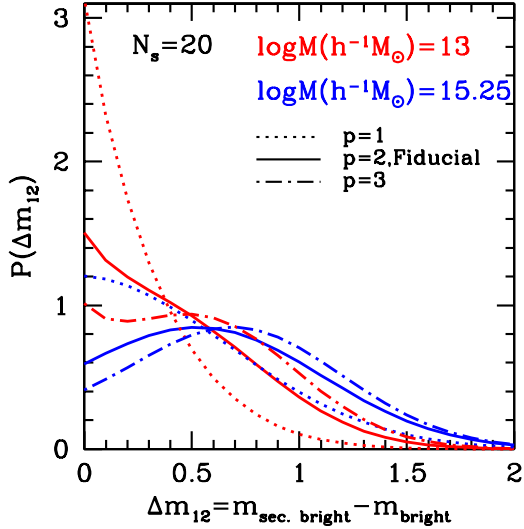} 
\caption{
The dependence of the distribution of magnitude gaps on the parameter
$p$ which governs the exponential cutoff at the bright end in the
conditional luminosity function of satellite galaxies, based on
Eq.~\ref{eq:pred}. The solid line corresponds to the fiducial case
$p=2$ while the dotted and the dot-dashed lines correspond to $p=1$
and $p=3$, respectively. The results for the two different halo masses
are shown by different colors.
}
\label{fig:varyp}
\end{figure}
First, we consider the effect of relaxing the assumption that the
centrals are the brightest in their halos. Note that in this case, the
magnitude gap could be either between the central and the brightest
satellite, or between the two brightest satellites, in case the halo
has two or more satellites brighter than the putative {\it central}
galaxy (see Appendix). The resultant magnitude gap is shown with a dashed histogram.
For low mass halos it can be hardly distinguished from the case when
we demand the central to be the brightest. It can be also seen that
for all halo masses the distribution of magnitude gaps for $\Delta
m_{12}>0.5$ is consistent with the case when the central galaxy is
assumed to be the brightest. This is also expected since the satellite
conditional luminosity function in our model dies exponentially at the
bright end.  Therefore if there is a satellite galaxy brighter than
the central galaxy, the magnitude gap is not expected to be extremely
large. Therefore, the few cases when the satellite galaxy is brighter,
cause a small but noticeable increase in the probability distribution
at the small magnitude gap end.

We show the results of varying the number of satellites in
Fig.~\ref{fig:maggap_varyn}. The left hand panel shows the magnitude
gap distribution in halos of different mass, when the number of
satellites equals 5, while the right hand panel shows the same for
number of satellites equal to 64. As the number of satellites
increases (decreases) the magnitude gap tends to be smaller (larger),
as expected (and qualitatively consistent with the results presented
in fig.~5 of \citealt{vdb2007}). The analytical expectation (from Eq.~\ref{eq:pred}) is
shown as a solid curve; it describes the simulation results
accurately, and is shown as a sanity check.

The parameter $p$ governs the exponential cut-off at the bright end of
the satellite conditional luminosity function (see Eq.~\ref{phi_s}).
Based upon the analysis of offsets of the line-of-sight velocities and
projected position of the brightest galaxy relative to the mean of the
other group members, \citet{Skibba2011} concluded that the value of
$p$ ought to be closer to unity instead of the fiducial value of $2$
that we assume \citep[see also][]{Reddick2012}. Therefore, we also
show the effect of varying the parameter $p$ on the magnitude gap
distribution in Fig.~\ref{fig:varyp}. We have verified that the
predictions based upon Eq.~\ref{eq:pred} that we show in the figure
also agree with detailed simulations. As expected, decreasing the
value of $p$ causes the satellite conditional luminosity function to
fall less rapidly at the bright end which results in smaller magnitude
gaps. 

Regardless of these details, it is clear that the results from this
section establish that if galaxies populate halos according to the
CLF, then at fixed richness the magnitude gap distribution should
depend upon the halo mass, in a manner which is qualitatively
consistent with \citet{Hearin2012}.

\section{Luminosity distribution of the brightest and second brightest galaxy}
\label{sec:lumdist}

\begin{figure*}
\centering
\includegraphics[scale=0.6]{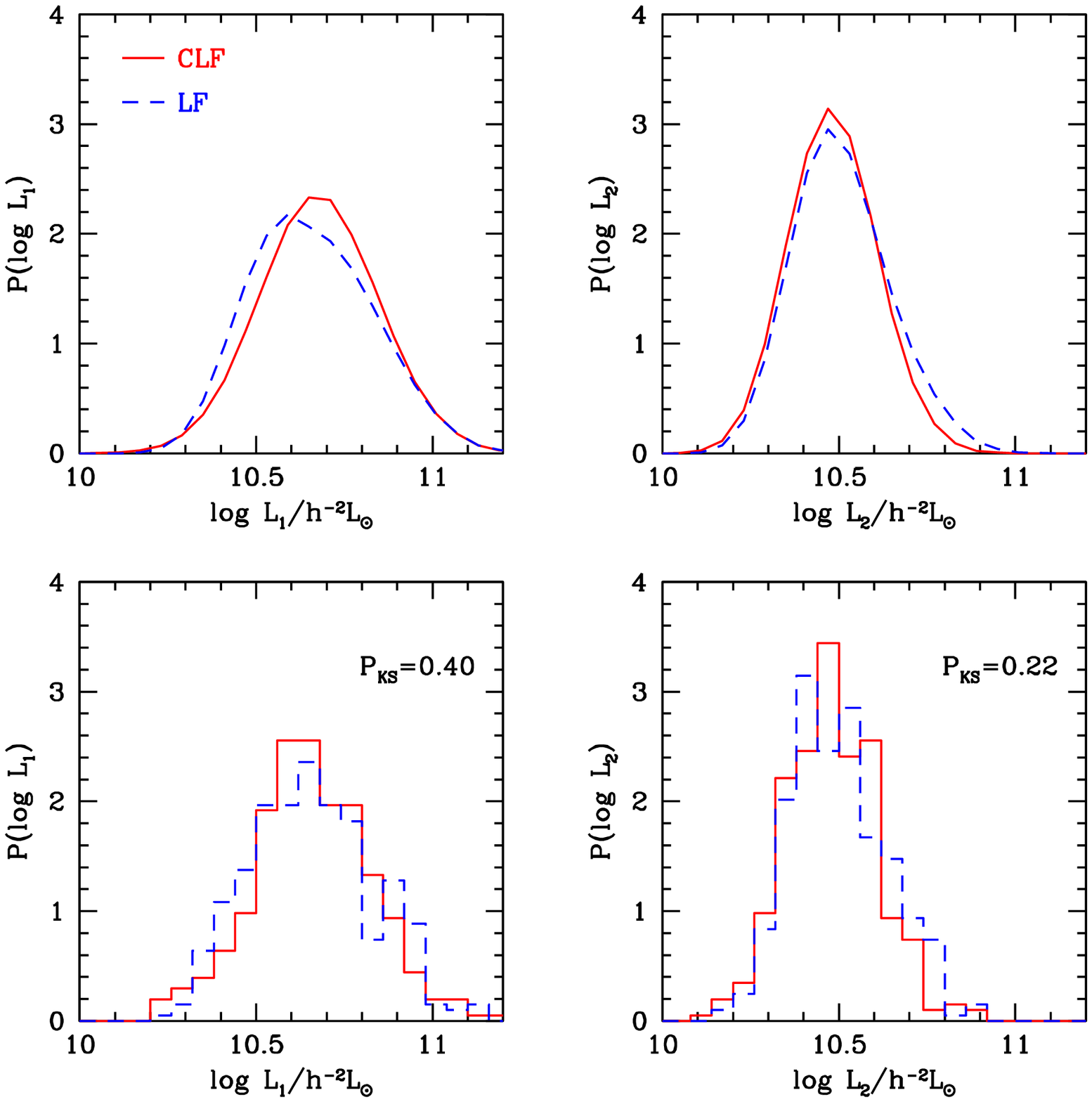} 
\caption{ Comparison between the luminosity distribution of the
brightest and the second brightest galaxy in the halos present in the
two mock galaxy catalogs are shown in the left and right hand side
panels, respectively. {\it Upper panels:} The solid line shows the luminosity
distribution when galaxies are populated in halos according to random
draws from the CLF (Catalog A), while the dashed histogram shows the
distribution when galaxies are populated according to random draws
from the global luminosity function (Catalog B), maintaining the richness
of halos.  {\it Bottom panels:} Same as the upper panels but for a
catalog with sample size comparable to the one used by
\citet{Paranjape2012}. The differences in the distribution from the
two catalogs, as quantified by the p-values from the KS-test are
indicated in each panel.
}
\label{fig:clf_lf}
\vspace{0.7cm}
\end{figure*}

\begin{figure}
\centering
\includegraphics[scale=0.8]{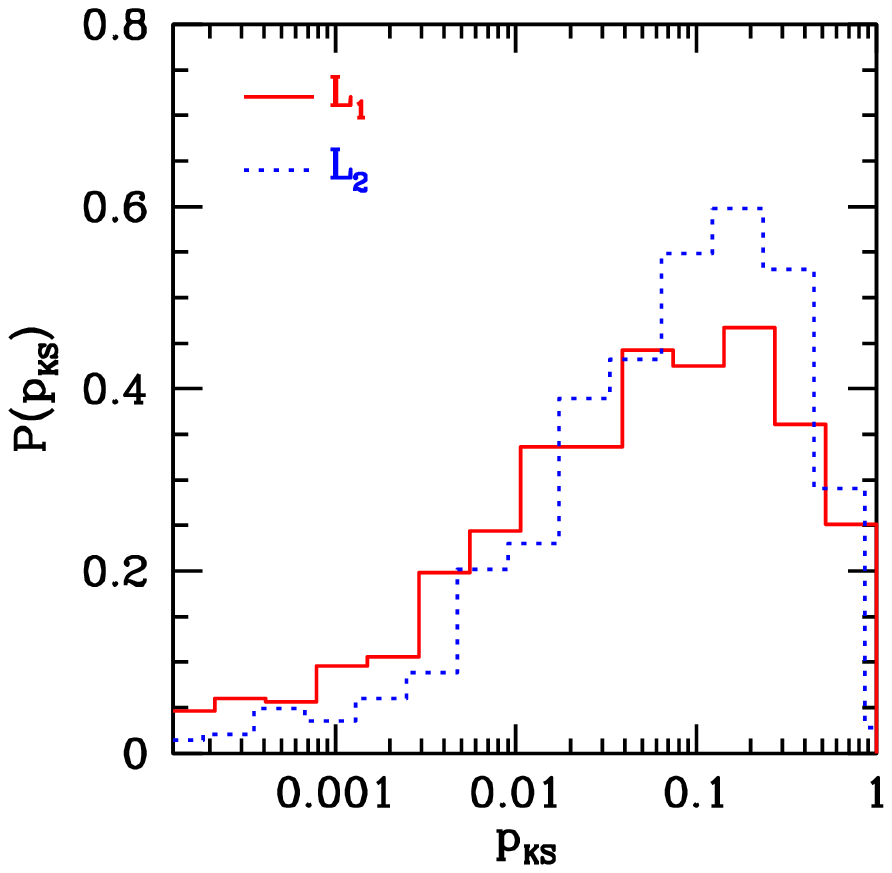} 
\caption{ Distribution of p-values from KS-test carried out on the
luminosity distribution of centrals (solid histogram) and satellites
(dotted histogram) from the two catalogs carried out on 1000 samples
with a sample size comparable to the one used by
\citet{Paranjape2012}.
}
\label{fig:ks}
\end{figure}

We would like to now
investigate the result presented in \citet{Paranjape2012}. They
demonstrate that the luminosity distribution of the brightest and the
second brightest galaxies in the group catalog of \citet{Berlind2006}
is consistent with their expected distribution if the luminosity of
galaxies in each of the groups were randomly sampled from the global
luminosity function of galaxies. To verify their result, we construct
Monte-Carlo galaxy catalogs in which galaxy luminosities are drawn
either from the conditional luminosity function or the overall
luminosity function.

We assume a standard flat $\Lambda$CDM cosmological model with matter
density $\Omega_\rmm = 0.27$, baryon density $\Omega_\rmb = 0.0469$,
Hubble parameter $h = 0.7$, spectral index $n_\rms=0.95$, and a matter
power spectrum normalization of $\sigma_8 = 0.82$. We sample a large
number of haloes with masses $M>2.7\times10^{13}\msunh$ from the halo mass
function expected for such cosmology using the halo mass function
calibration presented by \citet{Tinker2008}.

We construct a mock galaxy catalog (Catalog A) by populating the dark
matter halos with model galaxies using the CLF with parameters
described in \S\ref{sec:clf}. For each halo, we first draw the
luminosity of its central galaxy from $\Phi_{\rm cen}(L|M)$, given by
Eq.~(\ref{phi_c}).  Next, we draw the number of satellite galaxies,
under the assumption that $P(N_{\rm sat}|M)$ follows a Poisson
distribution with mean given by Eq.~(\ref{eq:ngal}) with $\Phi$
replaced by $\Phi_\rms$, and we adopt a luminosity threshold, $L_{\rm
min}$, corresponding to $^{0.1}M_r - 5\log h = -20$, similar to the
threshold adopted by \citet{Paranjape2012}. For each of the $N_{\rm
sat}$ satellites in the halo of question, we then draw a luminosity
from the satellite CLF $\Phi_{\rm sat}(L|M)$, given by
Eq.~(\ref{phi_s}) and maintain the fiducial assumption that all
satellites are fainter than the central galaxy. We restrict ourselves
to halos with richness $N\geq12$, which gives us a sample of 319482
halos.

We construct an alternate catalog of galaxies (Catalog B) where the
luminosities of member galaxies in each halo are drawn from the global
luminosity function, $\Phi(L)$,
\begin{equation}
    \Phi(L)=\int \Phi(L|M)n(M)\drm M\,,
\end{equation}
where $n(M)$ is the halo mass function. In practice, we randomly
sample (with replacement) from the luminosities of galaxies in
the entire previous catalog, while maintaining the richness of the
group they belong to, thus effectively sampling the galaxy
luminosities in every group from the global luminosity function of
galaxies.

The luminosity distribution of the brightest and the second brightest
galaxies in each halo for both the catalogs are shown in the upper
left and right hand panels of Fig.~\ref{fig:clf_lf}, respectively. The
peak of the magnitude distribution of the brightest galaxies in
Catalog A have a distribution which peaks at a slightly higher value
of luminosity compared to Catalog B. On the other hand, the magnitude
distribution of the second brightest galaxies shows a tail towards
larger luminosities in Catalog B compared to that in Catalog A.
However, the plot also shows that the differences are not that huge,
and detecting such differences in the magnitude distributions will
require a large sample of groups.

From our large sample of Monte-Carlo groups, we now restrict ourselves
to selecting sample sizes ($\sim350$) which are similar to those used
by \citet{Paranjape2012}. We show the results of one of the random
realizations in the bottom panel of Fig.~\ref{fig:clf_lf}. We also
obtain the corresponding cumulative distributions and use the
Kolmogorov-Smirnov (KS) statistic to compare the distributions from
the two catalogs. The p-values from the KS-test are indicated in the
corresponding panels and these values imply that the luminosity
distributions from the two catalogs, when downsampled to the size of
the catalog that \citet{Paranjape2012} use are consistent with each
other. To show that this particular random realization is not a
statistical fluke, in Fig.~\ref{fig:ks}, we show the distribution of
p-values from KS-tests carried out on 1000 random samples similar in
size to the catalog used by \citet{Paranjape2012}. The distribution of
p-values from the KS-test peak at values larger than 0.1, which
highlights the difficulty in distinguishing between the magnitude
distributions from the two catalogs with a small sample size.
This suggests that the group catalog used by \citet{Paranjape2012}
does not have enough number statistics, to detect the difference
between the luminosity distributions of the brightest (or the second
brightest) galaxies in the cases corresponding to the two catalogs. 

It is well known that the luminosity of central galaxies depends upon
the halo mass in which they reside. However, it is also known that at
the massive end, the luminosity of central galaxies is a weak function
of halo mass, e.g., based on two point statistics such as the
projected galaxy-galaxy correlation function, its dependence upon
luminosity of galaxies \citep{Zehavi2005,Zheng2007,Zehavi2011}, and
the projected galaxy-matter correlation function probed by the
galaxy-galaxy lensing measurements \citep{Mandelbaum2006,
Cacciato2009, Cacciato2012}, or other probes such as satellite
kinematics \citep{More2009a,More2011a} and subhalo abundance matching
\citep{Moster2010, Behroozi2010, Yang2012}.  This coupled with the
fact that the satellite fraction is very low at the bright end, could
be a reason why the differences in the magnitude distribution of the
brightest galaxy between the two different catalogs are not that
large.

We note that this insensitivity of the magnitude distributions to the
underlying halo occupation distribution was also pointed out by
\citet{Paranjape2012}, who suggested the use of two point statistics
such as the luminosity-marked correlation function, in order to
distinguish the two scenarios. Based on the radial dependence of the
marked correlation function, they concluded that the galaxy
luminosities in groups cannot be drawn from a global luminosity
function. However, their analysis does not directly address whether
this is due to a conditional luminosity function which varies with
mass, or a result of environmental dependences of the luminosity
function.



\section{Summary}
\label{sec:summary}

Recently, \citet{Hearin2012}, suggested that the magnitude gap between
the two brightest galaxies in a given halo at fixed richness contains
additional information about the halo mass. Their claim was based upon
an analysis of the galaxy group catalog constructed from the SDSS by
\citet{Berlind2006}. If correct, the magnitude gap information can be
used to reduce the scatter in the mass-richness relation in galaxy
clusters, which is important for the use of optically identified
galaxy clusters as cosmological probes. However, they claimed that
their result is at odds with the results presented in
\citet{Paranjape2012} who investigated the distribution of magnitudes
of the brightest and second brightest galaxies, from the same group
catalog. \citet{Paranjape2012} showed that these magnitude
distributions are consistent with the order statistics of the
luminosities sampled from the overall galaxy luminosity function
independent of halo mass. This would imply that the magnitude gap just
depends upon richness and does not contain extra information about the
halo mass.

We have investigated both these studies within the framework of the
conditional luminosity function (CLF), which describes the halo
occupation statistics of galaxies as a function of halo mass. The CLF
and its variation with halo mass has been calibrated using
observations of the abundance and clustering of galaxies, and the
galaxy-galaxy lensing signal in the SDSS, and is consistent with
results based upon the kinematics of satellite galaxies and abundance
matching. We have shown that if galaxies populate halos according to
the CLF and if the luminosities of central and satellite galaxies are
drawn from their corresponding CLF in an uncorrelated manner, then the
magnitude gap is expected to contain information about halo mass at
fixed richness. We have presented analytical expressions for
predicting the magnitude gap distribution at fixed richness as a
function of halo mass and verified these expressions using Monte-Carlo
simulation of galaxy catalogs populated according to the CLF.

We have shown that the magnitude distribution of the brightest
and the second brightest galaxies show significant differences,
between mock galaxy catalogs constructed by drawing galaxy
luminosities according to the CLF and those constructed according to
the luminosity function of galaxies. However, we have also shown that
these differences cannot be meaningfully detected given the small
sample size that \citet{Paranjape2012} use in their study. This shows
that the magnitude distribution of the brightest and the second
brightest galaxies is not the appropriate statistic to address the
issue of how galaxies populate dark matter halos, at least given the
current sample sizes.

These results suggest that the apparent tension between the two
studies is due to small sample size used by \citet{Paranjape2012}. The
magnitude gap at fixed richness can and does contain extra information
about the mass of a halo. As the sample size of groups grows, even the
luminosity distribution of the brightest and the second brightest
galaxies will also be able to distinguish between the two scenarios.
In this paper, we have provided an analytical model based on the CLF
to predict the magnitude gap distribution. We have also presented how
the magnitude gap distribution can vary as some of our fiducial
assumptions are changed.

It is also important to note that the CLF of galaxies in clusters can
also be directly observed, albeit as a function of optical properties
such as richness \citep[see e.g.,][]{Hansen2009}. Such observations,
when combined with halo mass indicators such as weak lensing, can in
turn be used to better constrain conditional luminosity function at
the high mass end, which will help to constrain our model for the
magnitude gap, at fixed richness.

Finally, we would like to remark that we have assumed that the
luminosities of the galaxies in every group are drawn from the
conditional luminosity function in an uncorrelated fashion. This
assumption, however, needs to be thoroughly tested. For example, if
bright satellite galaxies merge with the central galaxy due to
dynamical friction, the central galaxy will become brighter at fixed
halo mass, and the magnitude gap will correspondingly be larger.
However, simultaneously the richness of the group will decrease (which
will also cause the magnitude gap to be larger just due to the
statistics of random draws from the CLF), thus making it difficult to
disentangle the physical correlation from the effect due to changing
richness.


\section*{Acknowledgments}

I thank Andrew Hearin for discussion of his preliminary results during
my recent visit to the McWilliams Center for Cosmology at the Carnegie
Mellon University. I also thank Nick Gnedin, Andrew Hearin, Frank van
den Bosch and Andrew Zentner for their suggestions and useful comments
on an early draft of this manuscript. This research was supported by
the National Science Foundation under Grant No. NSF PHY-0551142 and a
generous endowment from the Kavli foundation.


\bibliographystyle{apj}
\bibliography{halos}

\appendix

\section{Magnitude gap when satellites are allowed to be brighter than
their central galaxies}

Based upon the analysis of offsets of the line-of-sight velocities and
projected position of the brightest galaxy relative to the mean of the
other group members, \citet{Skibba2011} concluded that there is a
significant chance that in a halo of given mass, a satellite galaxy is
brighter than the central galaxy. In Fig.~\ref{fig:maggap} presented
in Section~\ref{sec:sims}, we showed the magnitude gap distribution
for the case when we allow satellite galaxies to be brighter than
centrals. In this appendix, we provide analytical expressions for the
magnitude gap (between the two brightest galaxies) distribution in
this case.

Note that since we allow for the possibility that the satellite
galaxies can be brighter than the central galaxy, in a given halo, one
of the following three mutually exclusive and collectively exhaustive
cases may occur: (i) central galaxy still turns out to be the
brightest, (ii) central galaxy turns out to be second brightest, (iii)
central galaxy is not one of the two brightest galaxies.  The
probabilities corresponding to these three cases are given by
\begin{eqnarray}
P_1(M,N) &=& \int P_\rmc(L|M) \left[ P_\rms(<L|M) \right]^{N-1} \drm
L\,, \\
P_2(M,N) &=& (N-1)\int P_\rmc(L|M) [1-P_\rms(<L|M)] \left[
P_\rms(<L|M) \right]^{N-1} \drm L \,,\\
P_3(M,N)&=&1-P_1(M,N)-P_2(M,N) \,,
\end{eqnarray}
respectively. The magnitude gap distribution can then be expressed as
\begin{eqnarray}
P(f_L|M,N)&=&P_1(M,N)\,(N-1)\int_{L_{\rm min}}^{\infty} \drm L'\,
\,P_\rms(L'|M)\,P_\rmc\left(L'/f_L|M\right)\,\left[ P_\rms(<L'|M)
\right]^{(N-2)}\, \nonumber\\
&& + P_2(M,N)\,(N-1)\int_{L_{\rm min}}^{\infty} \drm L'\,
\,P_\rms(L'|M)\,P_\rmc\left(L'f_L|M\right)\,\left[
P_\rms(<L'f_L|M)\right]^{(N-2)} \nonumber \\
&& + P_3(M,N)\,(N-1)\,(N-2) \int_{L_{\rm min}}^{\infty} \drm L'\,
P_\rms(L'|M)\,P_\rms(L'/f_L|M)[P_\rms(<L'|M)]^{(N-3)}\,.
\end{eqnarray}

\end{document}